# Estimating regional ground-level PM$_{2.5}$ directly from satellite top-of-atmosphere reflectance using deep learning

Huanfeng Shen [a,b,c,*], Tongwen Li [a], Qiangqiang Yuan [d,b], Liangpei Zhang [e,b]

[a] School of Resource and Environmental Sciences, Wuhan University, Wuhan, China.

[b] The Collaborative Innovation Center for Geospatial Technology, Wuhan, China.

[c] The Key Laboratory of Geographic Information System, Ministry of Education, Wuhan University, Wuhan, China.

[d] School of Geodesy and Geomatics, Wuhan University, Wuhan, China.

[e] The State Key Laboratory of Information Engineering in Surveying, Mapping and Remote Sensing, Wuhan University, Wuhan, China.

[*] Corresponding author: Huanfeng Shen (shenhf@whu.edu.cn)

**Highlights**:

- We estimate ground PM$_{2.5}$ from satellite TOA reflectance rather than satellite AOD.
- The Ref-PM modeling achieves an outstanding performance for estimating ground PM$_{2.5}$.
- Daily fine-scale distributions of PM$_{2.5}$ are mapped in Wuhan Urban Agglomeration.

# ABSTRACT

Almost all remote sensing atmospheric PM$_{2.5}$ estimation methods need satellite aerosol optical depth (AOD) products, which are often retrieved from top-of-atmosphere (TOA) reflectance via an atmospheric radiative transfer model. Then, is it possible to estimate ground-level PM$_{2.5}$ directly from satellite TOA reflectance without a physical model? In this study, this challenging work are achieved based on a machine learning model. Specifically, we establish the relationship between PM$_{2.5}$, satellite TOA reflectance, observation angles, and meteorological factors in a deep learning architecture (denoted as Ref-PM modeling). Taking the Wuhan Urban Agglomeration (WUA) as a case study, the results demonstrate that compared with the AOD-PM modeling, the Ref-PM modeling obtains a competitive performance, with out-of-sample cross-validated R$^2$ and RMSE values of 0.87 and 9.89 $\mu g/m^3$ respectively. Also, the TOA-reflectance-derived PM$_{2.5}$ have a finer resolution and larger spatial coverage than the AOD-derived PM$_{2.5}$. This work updates the traditional cognition of remote sensing PM$_{2.5}$ estimation and has the potential to promote the application in atmospheric environmental monitoring.

# 1. Introduction

Fine particular matter (PM$_{2.5}$, airborne particles less than 2.5 $\mu m$ in the aerodynamic diameter) has been reported to be associated with many adverse health effects including cardiovascular and respiratory morbidity and mortality (Habre et al., 2014; Madrigano et al., 2013). Previous studies have indicated that the severe PM$_{2.5}$ pollution resulted in more than 3 million premature deaths around the world in the year of 2010 (Lim et al., 2012). It is thus an urgent need to acquire accurate spatiotemporal distributions of ground-level PM$_{2.5}$ concentration for the environmental health concerns.

The satellite-derived aerosol optical depth (AOD) products have been extensively employed to expand PM$_{2.5}$ estimation beyond that only provided by ground monitoring stations (Hoff and Christopher 2009; Li et al., 2016; van Donkelaar et al., 2016). The AOD products used include those retrieved from the Moderate Resolution Imaging Spectroradiometer (MODIS) (Fang et al., 2016), the Multiangle Imaging SpectroRadiometer (MISR) (You et al., 2015), the Geostationary Operational Environmental Satellite Aerosol/Smoke Product (GASP) (Paciorek et al., 2008), the Visible Infrared Imaging Radiometer Suite (VIIRS) (Wu et al., 2016), etc. In addition, many models were developed to establish the relationship between AOD and PM$_{2.5}$ (denoted as AOD-PM modeling), such as multiple linear regression (MLR) (Gupta and Christopher 2009b), geographically weighted regression (GWR) (Hu et al., 2013), linear mixed effects (LME) model (Lee et al., 2011), neural networks (NN) (Gupta and Christopher 2009a; Li et al., 2017b), and so on. Based on these models, the satellite-derived AOD products have played an important role in the estimation of ground-level PM$_{2.5}$.

The AOD products are often retrieved from satellite top-of-atmosphere (TOA) reflectance through an atmospheric radiative transfer model (e.g., 6S, MODTRAN) (Hsu et al., 2004; Kaufman et al., 1997; Levy et al., 2007). Hence, the procedure of previous satellite-based $PM_{2.5}$ estimation is usually to retrieve AOD from satellite TOA reflectance firstly, and subsequently estimate ground-level $PM_{2.5}$ from satellite-derived AOD. A challenging proposition is that whether it is possible to avoid the intermediate process of AOD retrieval, and to estimate ground-level $PM_{2.5}$ directly from satellite TOA reflectance (denoted as Ref-PM modeling). Actually, previous studies have provided some potentials for the Ref-PM modeling. On the one hand, some researchers (Radosavljevic et al., 2007; Ristovski et al., 2012) adopted neural networks to learn a functional relationship between MODIS observations and ground-observed AOD, and the neural-network-based AOD retrievals achieved satisfactory results compared with the physically retrieved AOD in their experiments. On the other hand, many aforementioned models have been employed to establish the AOD-$PM_{2.5}$ relationship. Given that the physical retrieval of AOD from TOA reflectance can be simulated using statistical approaches, as well as the AOD-$PM_{2.5}$ relationship, it appears possible to directly model the statistical relationship between TOA reflectance and ground-level $PM_{2.5}$.

Anyhow, it would be very complicated to estimate ground-level $PM_{2.5}$ from satellite TOA reflectance in one step. The retrieval of AOD from satellite TOA reflectance is a nonlinear physical problem; the satellite-derived AOD in conjunction with auxiliary factors (e.g., meteorological parameters) are also usually nonlinearly correlated with $PM_{2.5}$. Hence, the estimation of ground-level $PM_{2.5}$ directly from TOA reflectance is of high complexity, and the conventional models may encounter some challenges. Deep learning, which is the further development of neural network, has been used for time-series predictions (Ong et al., 2015) and satellite AOD-based estimation (Li et al., 2017a) of ground-level $PM_{2.5}$. The deep learning models have shown the capacity to effectively predict/estimate ground-level $PM_{2.5}$, which can be attributed to its capacity of fitting nonlinear and complicated relationship (Hinton et al., 2006). Thus, deep learning may be a good tool for the satellite TOA reflectance-based estimation of ground-level $PM_{2.5}$.

The objective of this study is to develop a deep learning-based modeling for the estimation of ground-level $PM_{2.5}$ using satellite TOA reflectance rather than AOD products. To be specific, one of the most typical deep learning models (i.e., deep belief network) will be employed to establish the relationship between ground-level $PM_{2.5}$, satellite TOA reflectance, observation angles, and meteorological factors. Through the Ref-PM modeling, we can not only simplify the procedure of satellite-based $PM_{2.5}$ estimation, but also avert the accumulative error of AOD retrieval, which has been reported in previous studies (Munchak et al., 2013). The proposed Ref-PM modeling will be validated with the data from Wuhan Urban Agglomeration (Fig. 1) in the whole year of 2016.

**2. Study region and data**

*2.1. Study region*

The study region is Wuhan Urban Agglomeration (WUA), which is presented in Fig. 1. The study period is a total year of 2016. WUA is located in Hubei province, central China (as shown in Fig. 1 (a) and (b)). To make full use of $PM_{2.5}$ station measurements, the monitoring stations in the range with latitude of 28.4 °~32.3 °N and longitude of 112.0 °~116.7 °E are all adopted in our analysis. WUA is a city group with the center of Wuhan, covering the vicinal 8 cities (Huangshi, Ezhou, Huanggang, Xiaogan, Xianning, Xiantao, Qianjiang, and Tianmen). It has the total population of greater than 30 million, which account for more than a half of the total population of Hubei province. With over 60% of Gross Domestic Product (GDP) of Hubei province, WUA is one of the largest urban groups in central China. In 2007, it is authorized as one of the reforming pilot areas for national resource-saving and environment-friendly society by the Chinese government.

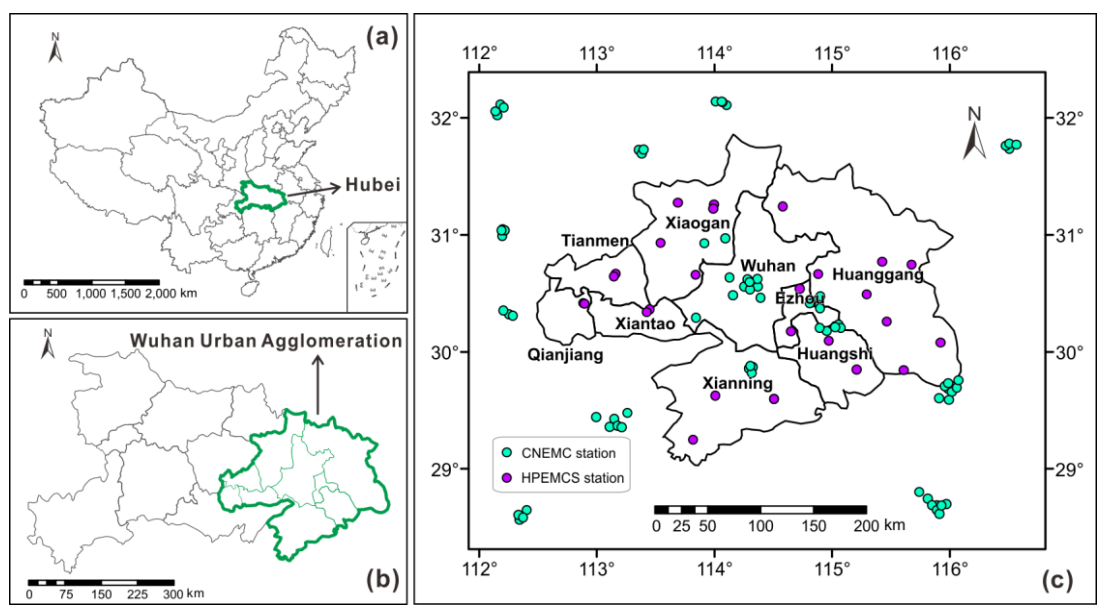

**Fig. 1**. Study region and the spatial distribution of PM$_{2.5}$ stations. (a) The location of Hubei province in China. (b) The boundary of Wuhan Urban Agglomeration in Hubei province. (c) The Wuhan Urban Agglomeration and the distribution of PM$_{2.5}$ monitoring stations.

Due to the dense urbanization and industrial activities, WUA has been suffering serious air pollution. Wuhan is especially populated with fine-mode particles (Wang et al., 2015), and the mean atmospheric PM$_{2.5}$ mass concentration was about $160 \pm 50\ \mu g/m^3$ (Wang et al., 2014). Moreover, with the rapid economic development, the other cities in WUA (e.g., Ezhou and Huangshi) also have a high level of PM$_{2.5}$ concentration. It is thus very urgent and necessary to obtain the spatiotemporal distributions of PM$_{2.5}$ in this area.

*2.2. Ground-level PM$_{2.5}$ measurements*

Hourly PM$_{2.5}$ concentration data within the study region in 2016 were obtained from the China National Environmental Monitoring Center (CNEMC) website (http://www.cnemc.cn) and the Hubei Provincial Environmental Monitoring Center Station (HPEMCS) website (http://www.hbemc.com.cn/). In this study, 77 CNEMC stations and 27 HPEMCS stations (104 stations in total) are included. The distribution of PM$_{2.5}$ monitoring stations is shown in Fig. 1 (c). It can be found that the CNEMC stations are unevenly distributed, while the HPEMCS stations make a good compensation. According to the Chinese National Ambient Air Quality Standard (CNAAQS, GB3905-2012), the ground PM$_{2.5}$ concentrations are measured by the tapered element oscillating microbalance method (TEOM) or with beta attenuation monitors (BAMs or beta-gauge), with an uncertainty of 0.75% for the hourly record (Engel-Cox et al., 2013). We averaged hourly PM$_{2.5}$ to daily mean PM$_{2.5}$ data for the satellite-based estimation of PM$_{2.5}$. For each monitoring station, the dates with less than 18 hourly observations were excluded in our analysis.

*2.3. Satellite observations*

The Aqua Moderate Resolution Imaging Spectroradiometer (MODIS) Level 1B calibrated radiances (MYD02) product was downloaded from the Level 1 and Atmosphere Archive and Distribution System (LAADS) website (https://ladsweb.modaps.eosdis.nasa.gov). They have a spatial resolution of 1 km at nadir. The top-of-atmosphere reflectance on bands 1, 3 and 7 (R1, R3 and R7), and observation angels (i.e., sensor azimuth, sensor zenith, solar azimuth and solar zenith), were exploited for the retrieval of AOD via dark-target-based algorithms (Kaufman et al., 1997). Despite of the avoidance of AOD retrieval, these parameters were still extracted from this product to estimate ground-level PM$_{2.5}$. To eliminate the cloud contamination, the MODIS cloud mask product (MYD35_L2) was adopted. They are available at a resolution of 1 km every day. The cloud mask products have four confidence levels, i.e., "cloudy", "uncertain clear", "probably clear", and "confident clear" (Ackerman et al., 1998). In this study, we only adopted the data with the highest confidence level ("confident clear").

Furthermore, the MODIS normalized difference vegetation index (NDVI) product (Level 3, MYD13) with a resolution of 1 km every 16 days, were also downloaded from the LAADS website. The MODIS NDVI was incorporated into the PM$_{2.5}$ estimation model to reflect the land cover type. For a comparison purpose, the MODIS aerosol optical depth (AOD) products of Collections 6 were utilized to establish the AOD-PM modeling. They have a spatial resolution of 3 km. The 3-km AOD products were retrieved using the dark-target-based algorithm (Munchak et al., 2013).

*2.4. Meteorological data*

The level of PM$_{2.5}$ concentration is associated with meteorological parameters (Yang et al., 2017), the Goddard Earth Observing System Data Assimilation System GEOS-5 Forward Processing (GEOS 5-FP) (Lucchesi 2013) meteorological data were incorporated in this study. GEOS 5-FP exploits an analysis developed jointly with NOAA's National Centers for Environmental Prediction (NCEP), which allows the Global Modeling and Assimilation Office (GMAO) to take advantage of the developments at NCEP and the Joint Center for Satellite Data Assimilation (JCSDA). The reanalysis meteorological data have a spatial resolution of 0.25 ° latitude × 0.3125 ° longitude. We extracted relative humidity (RH, %), air temperature at a 2 m height (TMP, K), wind speed at 10 m above ground (WS, m/s), surface pressure (PS, kPa), and planetary boundary layer height (PBL, m) between 1 p.m. and 2 p.m. local time (the Aqua satellite overpass time corresponds to 1:30 p.m. local time). More details can be found at its official website (https://gmao.gsfc.nasa.gov/forecasts/).

*2.5. Data pre-processing and matching*

Firstly, we created a 0.01-degree grid and a 0.03-degree grid for the Ref-PM modeling and the AOD-PM modeling, respectively. The data matching, model establishment, and spatial mapping are based on the established grids. For each grid, ground-level $PM_{2.5}$ measurements from multiple stations are averaged. Meanwhile, we resampled the meteorological data to match with satellite observations. All these data were re-projected to the same coordinate system. Finally, we extracted satellite observations, meteorological parameters on the locations where the $PM_{2.5}$ measurements are available.

## 3. Deep learning-based Ref-PM modeling for the estimation of $PM_{2.5}$

In the process of AOD retrieval, the satellite TOA reflectance bands and observation angles are utilized as the primary input. Also, an atmospheric radiative transfer model is exploited to simulate the physical relationship between reflectance, aerosol mode, and observation geometry. Though the Ref-PM modeling is aimed to avoid the complicated process of AOD retrieval, and to estimate ground-level $PM_{2.5}$ directly from satellite TOA reflectance, the original input (i.e., TOA reflectance and observation angels) are still adopted for the estimation of ground-level $PM_{2.5}$.

Generally, the structure of the Ref-PM modeling is depicted as Eq. (1). The dependent variable is $PM_{2.5}$ concentration; and the input predictors are satellite TOA reflectance, observation angles, meteorological parameters, and satellite NDVI. Additionally, we incorporated the geographical correlation of PM$_{2.5}$ into the Ref-PM modeling. Because the nearby PM$_{2.5}$ from neighboring $s$ grids and the PM$_{2.5}$ observations from prior $t$ days for the same grid are informative for estimating PM$_{2.5}$. Here, $s, t$ are set as 5 and 3 respectively.

$$PM_{2.5} = f\left(R1, R3, R7, angles, RH, WS, TMP, PBL, PS, NDVI, S\text{-}PM_{2.5}, T\text{-}PM_{2.5}, DIS\right) \quad (1)$$

where $S\text{-}PM_{2.5}, T\text{-}PM_{2.5}, DIS$ denotes the geographical correlation of PM$_{2.5}$, and their calculations can refer to our previous study (Li et al., 2017a). R7, which is slightly influenced by atmosphere, reflects the surface reflectance. R1 and R3 can reflect the surface-atmosphere coupled reflectance on red and blue channels, and the surface reflectance can be separated using an empirical relationship (e.g., dark target algorithm (Kaufman et al., 1997)). Hence, the impact of atmosphere on satellite TOA reflectance signals can be separated from these input variables, and this is the basic principle and key point for remote sensing of PM$_{2.5}$.

The relationship between PM$_{2.5}$, satellite TOA reflectance, observation angles and meteorological factors is very complicated. Thus, deep learning, which has great potential for fitting the nonlinear and complex relationship, was employed to represent this relationship. To be specific, one of the most typical deep learning models (i.e., deep belief network, DBN) (Hinton et al., 2006) was adopted. Fig. 2 presents the structure of a DBN model containing two hidden layers. As illustrated in the figure, the basic unit is a restricted Boltzmann machine (RBM). An RBM contains a visible layer and a hidden layer, where the hidden layer of the prior RBM is the visible layer of the next RBM. The DBN consists of multiple RBMs and a back-propagation (BP) layer. This BP layer can be utilized for classification or prediction, and it is here used for the prediction of ground-level PM$_{2.5}$. In this study, two RBM layers with 15 neurons in each RBM layer are adopted. The details of the DBN model can refer to Li et al., (2017a). The procedure of this model for the Ref-PM modeling is divided to three steps, which is shown in Fig.3.

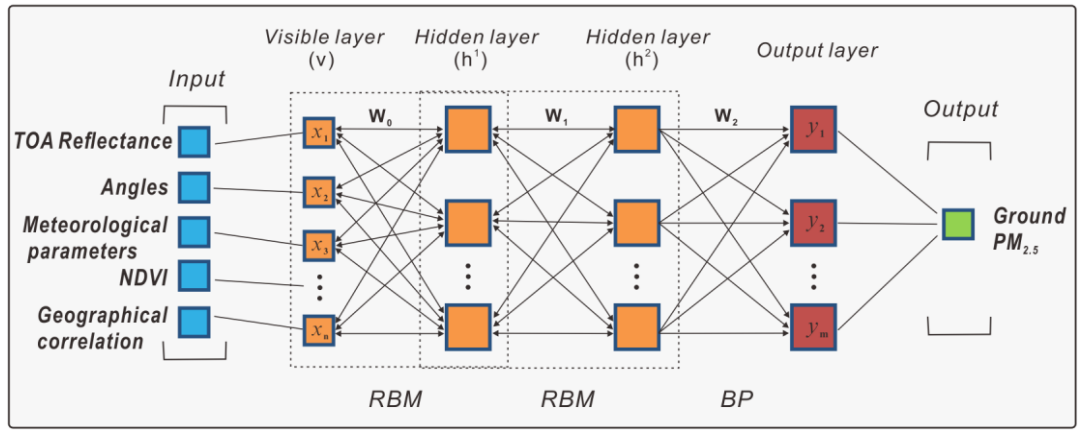

**Fig. 2**. The structure of DBN model for the Ref-PM modeling.

**Step 1**: The satellite TOA reflectance, observation angles, meteorological parameters, NDVI, and geographical correlation of $PM_{2.5}$ are input into the DBN model. This model is pre-trained without supervision to initialize itself. That is, the station $PM_{2.5}$ measurements are not used in this step, the initial model coefficients are trained from input data.

**Step 2**: The estimated $PM_{2.5}$ can be obtained from the DBN model. Subsequently, we calculate mean square error (MSE) between estimated $PM_{2.5}$ and ground observed $PM_{2.5}$. The error is sent back to fine-tune the model coefficients using the back-propagation (BP) algorithm (Rumelhart et al., 1986). This process will be repeated until the model achieves a satisfactory performance. Through this step, the DBN model can effectively establish the relationship between $PM_{2.5}$ and satellite reflectance.

**Step 3**: The model will be validated and exploited to predict the $PM_{2.5}$ values for those locations with no ground monitoring stations. Thus, the distribution of $PM_{2.5}$ concentrations can be acquired.

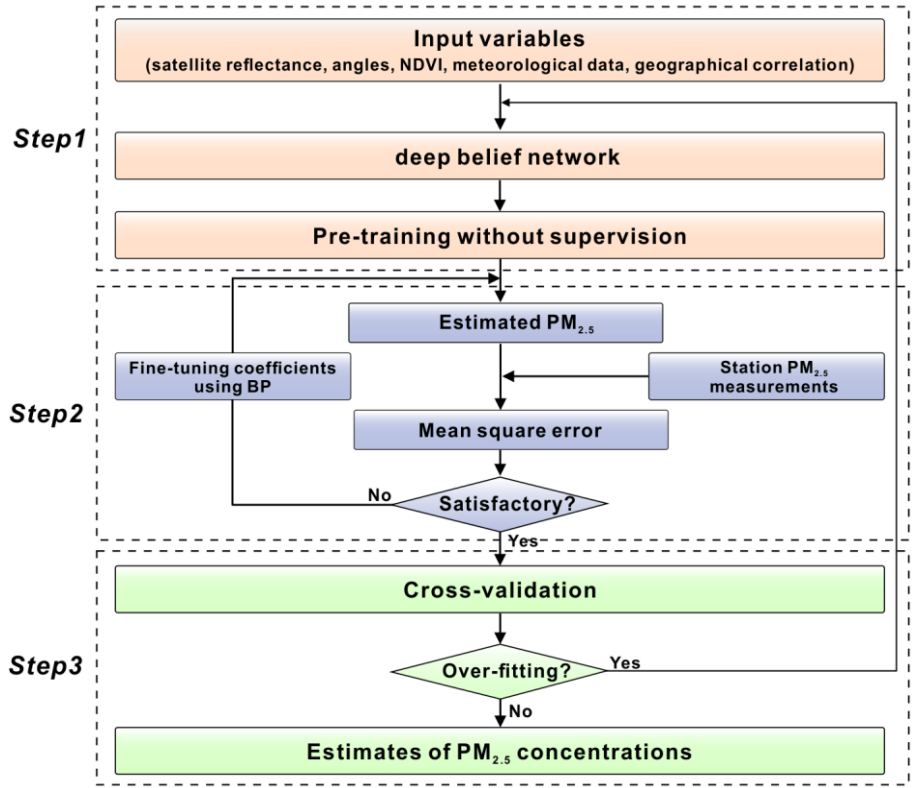

**Fig. 3**. Flowchart describing the process of DBN model for the Ref-PM modeling.

A 10-fold cross-validation (CV) technique (Rodriguez et al., 2010) was adopted to test the potential of model overfitting and predictive power. Previous studies usually used sample-based CV (Li et al., 2017b; Ma et al., 2014) or site-based CV (Lee et al., 2011; Xie et al., 2015) to evaluate the model performance. In this study, both sample-based CV and site-based CV were chosen for the model validation. For sample-based CV, all samples in the model dataset were randomly and equally divided into ten subsets. One subset was used as validation samples and the rest subsets were used to fit the model for each round of validation. For site-based CV, we divided the monitoring stations into ten subsets randomly and equally. One subset was used for validation and the remaining stations for model fitting in each round. We adopted the statistical indicators of the coefficient of determination ($R^2$), the root-mean-square error (RMSE, $\mu g/m^3$), the mean prediction error (MPE, $\mu g/m^3$), and the relative prediction error (RPE, defined as RMSE divided by the mean ground-level PM$_{2.5}$) to give a quantitative evaluation of the model performance.

## 4. Results and discussion

*4.1. Descriptive statistics*

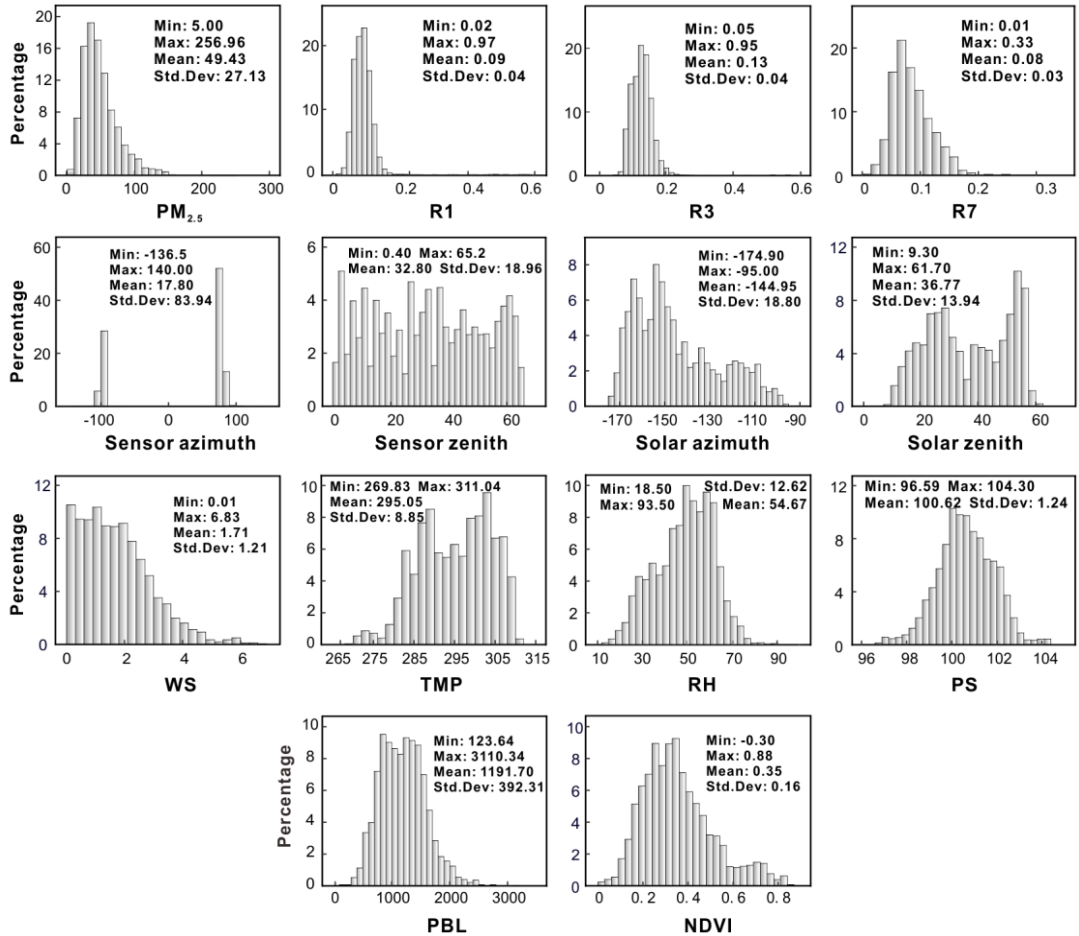

**Fig. 4**. Histograms and descriptive statistics of the Ref-PM modeling variables in the sample dataset.

Fig. 4 shows the histograms and descriptive statistics of the variables in the sample dataset. The PM$_{2.5}$ concentrations range from $5\,\mu g/m^3$ to $256.96\,\mu g/m^3$, with an average of $49.43\,\mu g/m^3$. The R1, R3, and R7 are mostly distributed in 0~0.2, with mean values of 0.09, 0.13, and 0.08, respectively. It can be found that satellite TOA reflectance, temperature, relative humidity, surface pressure, height of planetary boundary layer, and NDVI have similar distribution patterns with PM$_{2.5}$ concentration, while the observation angles and wind speed have different distributions. The correlation coefficient between R1 (R3) and PM$_{2.5}$ is 0.10 (0.12), indicating the nonlinear relationship between TOA reflectance and PM$_{2.5}$.

*4.2. Performance of the Ref-PM modeling*

Fig. 5 shows the cross-validation performance of the Ref-PM modeling and the AOD-PM modeling, respectively. The sample size of the AOD-PM modeling dataset (N=1658) is much smaller than the Ref-PM modeling dataset (N=4181). The main reason is that the TOA reflectance data has a finer spatial solution and larger coverage than the AOD product. For the Ref-PM modeling, an outstanding performance is achieved, with the sample-based (site-based) cross-validation $R^2$ and RMSE values of 0.87 (0.79) and 9.89 (12.97) $\mu g/m^3$, respectively. Meanwhile, the sample-based (site-based) cross-validation $R^2$ and RMSE of the AOD-PM modeling are 0.86 (0.72) and 10.42 (15.30) $\mu g/m^3$, respectively. We can observe that the sample-based cross-validation results of the Ref-PM modeling ($R^2$=0.87) merely shows a slight advantage than that of the AOD-PM modeling ($R^2$=0.86). Whilst the site-based cross-validation results of the AOD-PM modeling report a relatively larger decrease (from 0.79 to 0.72 for $R^2$) compared with that of the Ref-PM modeling. It is worth nothing that the site-based cross-validation approach, which uses a spatial hold-out validation strategy, can reflect the spatial predictive power more adequately (Li et al., 2017a). Hence, the results indicate that the Ref-PM modeling has a relatively superior spatial predictive power than the AOD-PM modeling.

On the other hand, the sample-based and site-based cross-validation slopes for the Ref-PM modeling are 0.87 and 0.82, respectively. This means that the proposed Ref-PM modeling tends to underestimate when the ground PM$_{2.5}$ concentrations are greater than ~50 $\mu g/m^3$.

Meanwhile, the AOD-PM modeling reports a little higher extent of underestimation, with the sample-based and site-based cross-validation slopes of 0.82 and 0.79 respectively. The possible reason for this underestimation could be that we exploited point-based monitoring data and a spatially averaged modeling framework. The sampling distribution of monitoring stations in a grid may not give a great estimation of the spatially averaged concentration for that grid (Li et al., 2017a).

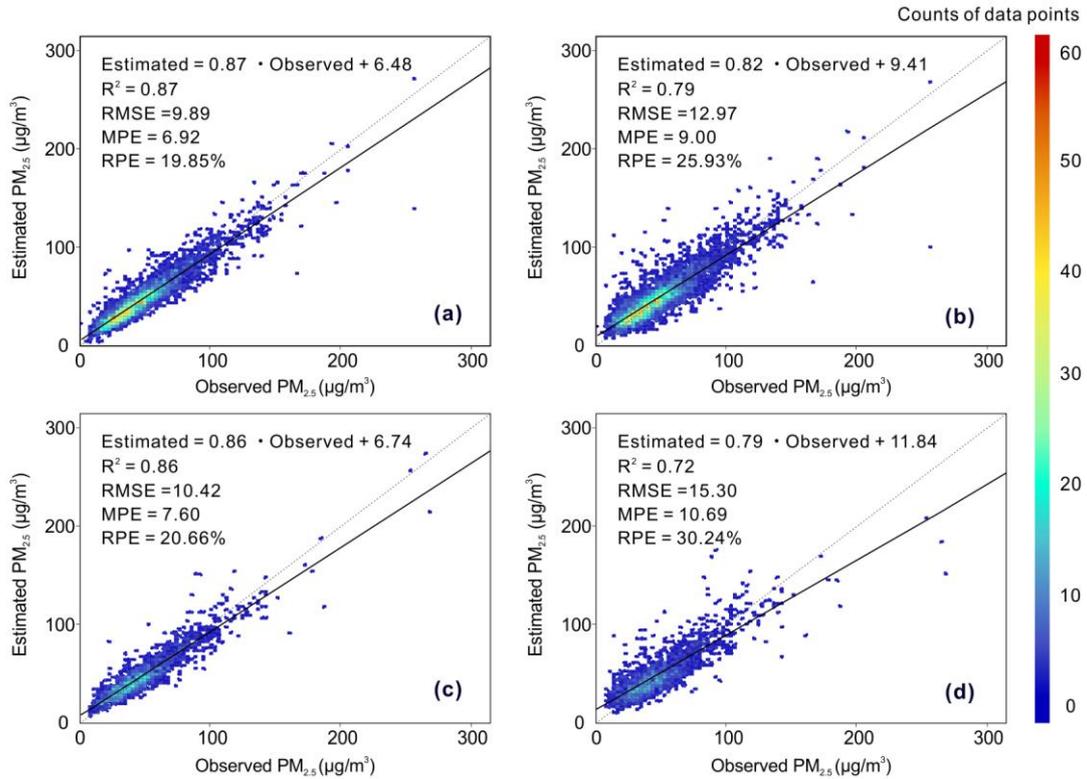

**Fig. 5.** Scatter plots of the Ref-PM modeling and the AOD-PM modeling cross-validation results: (a) and (b) are sample-based and site-based cross-validation results of the Ref-PM modeling (N=4181); (c) and (d) are sample-based and site-based cross-validation results of the AOD-PM modeling (N=1658).

Furthermore, the spatial site-based cross-validation performance of the Ref-PM modeling and the AOD-PM modeling was evaluated respectively. The $R^2$ and RMSE values between the observed $PM_{2.5}$ and estimated $PM_{2.5}$ on each grid were calculated, which are shown in Fig. 6. Overall, the Ref-PM modeling has achieved a satisfactory performance, with 70% of grids reporting a high $R^2$ value of greater than 0.80, and the low RMSE values ($<15\ \mu g/m^3$) are found on 81% of grids. For the AOD-PM modeling, merely 59% of grids have $R^2$ values of greater than 0.80, and the grids with RMSE values of less than 15 $\mu g/m^3$ account for 69%. These statistics indicate that the Ref-PM modeling shows some superiorities in the spatial prediction of $PM_{2.5}$ than the AOD-PM modeling. Also, some spatial variations are observed in Fig. 6. For instance, compared with the results of the AOD-PM modeling, an obvious advantage is found in Wuhan for the Ref-PM modeling. On the contrary, the AOD-PM modeling achieves a slightly better result than the Ref-PM modeling in the regions of Xianning and Huanggang.

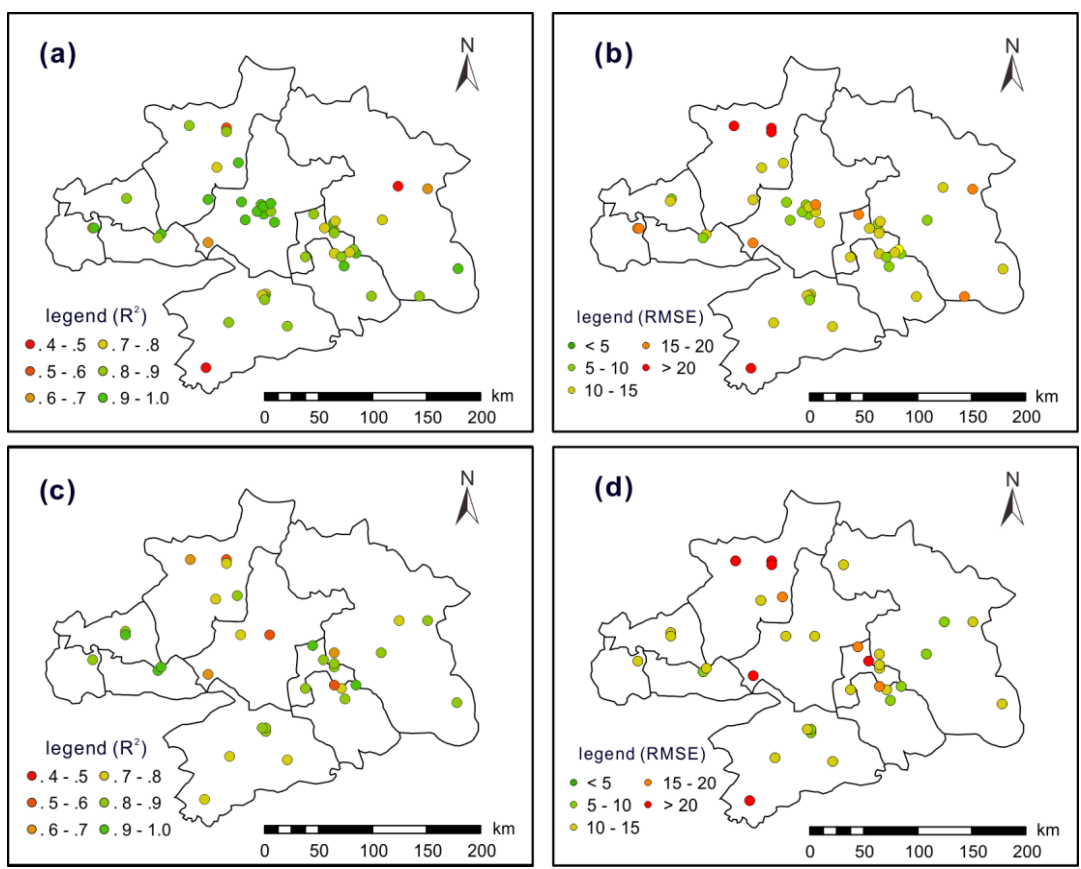

**Fig. 6.** Spatial performance of site-based cross-validation: (a) and (b) are $R^2$ and RMSE for the Ref-PM modeling respectively; (c) and (d) are $R^2$ and RMSE for the AOD-PM modeling respectively.

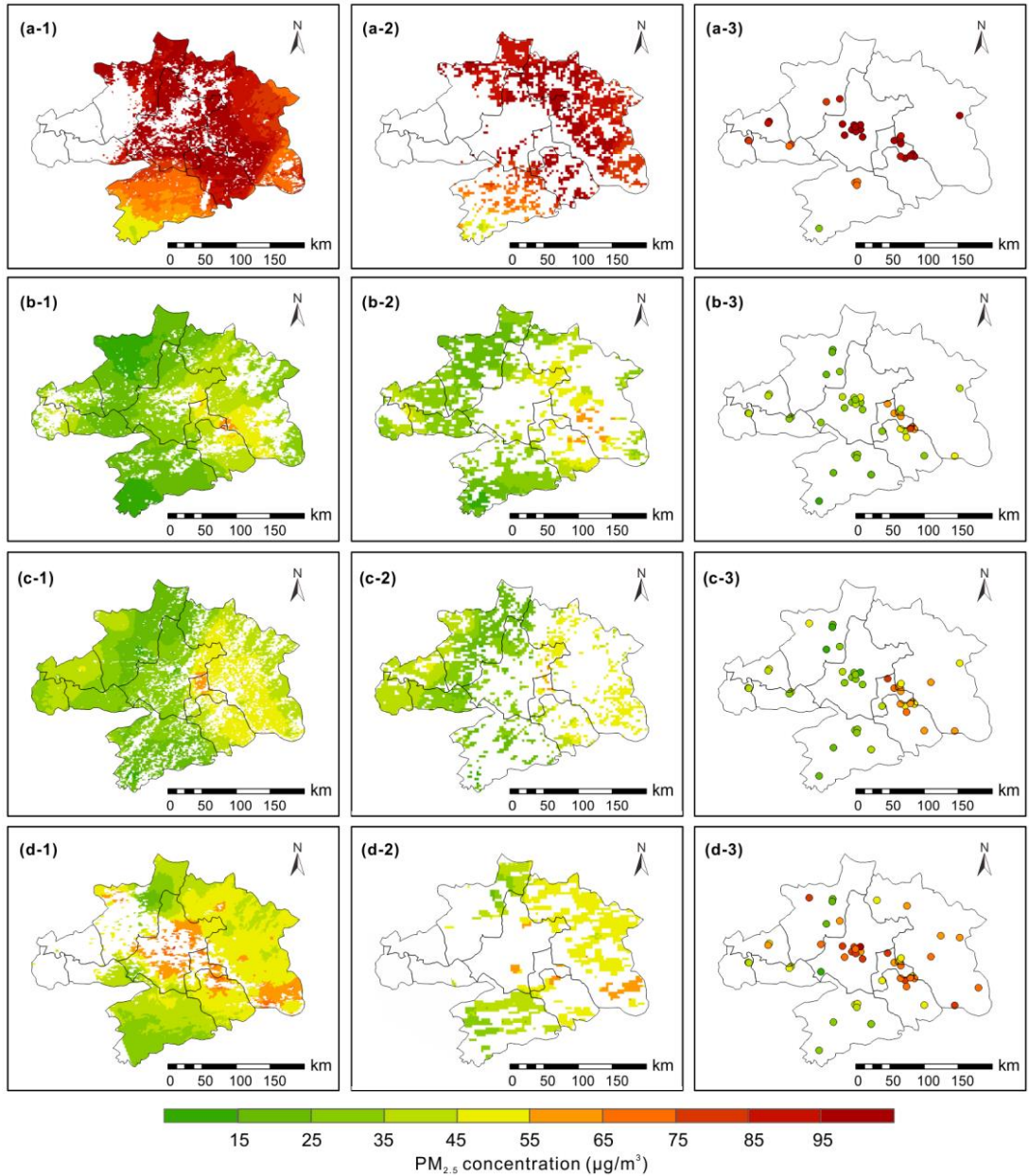

**Fig. 7**. Daily estimates of $PM_{2.5}$ on some specific days (one day in each season), which have as much as valid data. From top to bottom: (a) 20160115, (b) 20160516, (c) 20160725, and (d) 20161127. Left column: TOA reflectance-based estimation of $PM_{2.5}$. Middle column: AOD-based estimation of $PM_{2.5}$. Right column: ground station measurements. The white regions indicate missing data.

*4.3. Discussion*

The retrieval of AOD often meets a challenge, which is the bright surfaces (R7>0.15) (Li et al., 2005). Thus, the AOD-based estimation of $PM_{2.5}$ may be missing on the bright surfaces regions. In this paper, the TOA reflectance on bright surface have been included to estimate ground-level $PM_{2.5}$. As illustrated in Fig. 7, the reflectance-derived $PM_{2.5}$ have larger spatial coverage than the AOD-derived $PM_{2.5}$, which can largely be attributed to the estimates on bright surfaces. We evaluate the performance of the Ref-PM modeling on the bright surfaces. The results show that the sample-based cross-validated $R^2$ value is 0.80 on the bright surface regions, and 0.87 on the dark target regions. We can see that the accuracy of $PM_{2.5}$ estimation on bright surfaces is acceptable. Also, the reflectance-derived $PM_{2.5}$ have a smooth tendency in space (Fig. 7). The results indicate that the Ref-PM modeling has the potential to provide larger coverage of valid $PM_{2.5}$ estimates than the AOD-PM modeling. As is further observed in Fig. 7, the reflectance-derived $PM_{2.5}$ have a finer resolution than the AOD-derived $PM_{2.5}$, so that more details of $PM_{2.5}$ distributions are able to be investigated. These findings reveal that the reflectance-derived $PM_{2.5}$ have a finer resolution and more valid estimates on bright surfaces than the AOD-derived $PM_{2.5}$.

This study provides an effective solution for the estimation of ground-level $PM_{2.5}$ from satellite TOA reflectance rather than AOD products. However, the satellite-derived AOD is still one of the import atmospheric parameters. For example, the satellite-derived AOD itself is an indicator of atmospheric pollution, and plays an important role in the researches on aerosol-cloud interaction, climate changes, etc. This paper merely says that as for the estimation of ground-level $PM_{2.5}$, it appears possible to avoid the retrieval of satellite AOD. The proposed solution can also be extended to the monitoring of other environmental features, especially when no proper satellite parameters like AOD are found. Therefore, this study has no intention to belittle AOD retrieval or the AOD-PM modeling, but emphasizes on developing a new solution for the satellite-based estimation of $PM_{2.5}$.

## 5. Conclusions and future work

In this paper, the Ref-PM modeling is developed to avoid the intermediate process of AOD retrieval, and to estimate ground-level PM$_{2.5}$ directly from satellite TOA reflectance without a physical model. More importantly, this Ref-PM modeling has the potential to be extended to the satellite-based monitoring of other environmental features. Using the Wuhan Urban Agglomeration (WUA) as an example, the results show that compared with the AOD-PM modeling, the Ref-PM modeling achieves a competitive performance, with sample-based cross-validation $R^2$ and RMSE values of 0.87 and 9.89 $\mu g/m^3$ respectively. Also, the daily distributions of PM$_{2.5}$ in WUA are mapped, and the reflectance-derived PM$_{2.5}$ have a finer resolution and larger spatial coverage than the AOD-derived PM$_{2.5}$. All these results can say that the proposed Ref-PM modeling has the capacity to estimate ground-level PM$_{2.5}$ concentration directly from satellite TOA reflectance. This work updates the traditional cognition of remote sensing PM$_{2.5}$ estimation and can provide useful information for the pollution monitoring and control.

The future studies will focus on two aspects. Firstly, the Ref-PM modeling has achieved some reasonable results for the estimation of PM$_{2.5}$ directly from TOA reflectance. However, we only selected R1, R3, R7, and observation angles for the model establishment. The selection is inspired by dark target algorithm (Kaufman et al., 1997). Will the incorporation of surface reflectance data (e.g., MOD09) boost the accuracy of PM$_{2.5}$ estimation? Whether or not more/other satellite parameters can better explain PM$_{2.5}$ concentration still has room for researching. Secondly, the proposed Ref-PM modeling is validated in a small region of WUA in this study. It may bring out new challenges and problems in larger geographical regions.

The application and validation of the Ref-PM modeling in large geographical regions will be conducted in the future studies.

**Acknowledgments**

This work was funded by the National Key R&D Program of China (2016YFC0200900) and the National Natural Science Foundation of China (41422108). We are grateful to the China National Environmental Monitoring Center (CNEMC), the Hubei Provincial Environmental Monitoring Center Station (HPEMCS), the Goddard Space Flight Center Distributed Active Archive Center (GSFC DAAC), the US National Aeronautics and Space Administration (NASA) Data Center for providing the foundational data.

ensemble with adaptive cost function. In, *the 10th International Conference on Engineering Applications of Neural Networks* (pp. 266-275)

Ristovski, K., Vucetic, S., & Obradovic, Z. (2012). Uncertainty Analysis of Neural-Network-Based Aerosol Retrieval. *IEEE Transactions on Geoscience and Remote Sensing, 50*, 409-414

Rodriguez, J.D., Perez, A., & Lozano, J.A. (2010). Sensitivity Analysis of k-Fold Cross Validation in Prediction Error Estimation. *IEEE Transactions on Pattern Analysis and Machine Intelligence, 32*, 569-575

Rumelhart, D.E., Hinton, G.E., & Williams, R.J. (1986). Learning representations by back-propagating errors. *Nature, 323*, 533-536

Sorek-Hamer, M., Kloog, I., Koutrakis, P., Strawa, A.W., Chatfield, R., Cohen, A., Ridgway, W.L., & Broday, D.M. (2015). Assessment of PM2.5 concentrations over bright surfaces using MODIS satellite observations. *Remote Sensing of Environment, 163*, 180-185

van Donkelaar, A., Martin, R.V., Brauer, M., Hsu, N.C., Kahn, R.A., Levy, R.C., Lyapustin, A., Sayer, A.M., & Winker, D.M. (2016). Global Estimates of Fine Particulate Matter using a Combined Geophysical-Statistical Method with Information from Satellites, Models, and Monitors. *Environmental Science & Technology, 50*, 3762-3772

Wang, L., Gong, W., Li, J., Ma, Y., & Hu, B. (2014). Empirical studies of cloud effects on ultraviolet radiation in Central China. *International Journal of Climatology, 34*, 2218-2228

Wang, L., Gong, W., Xia, X., Zhu, J., Li, J., & Zhu, Z. (2015). Long-term observations of aerosol optical properties at Wuhan, an urban site in Central China. *Atmospheric Environment, 101*, 94-102

Wu, J., Yao, F., Li, W., & Si, M. (2016). VIIRS-based remote sensing estimation of ground-level PM2.5 concentrations in Beijing–Tianjin–Hebei: A spatiotemporal statistical model. *Remote Sensing of Environment, 184*, 316-328

Xie, Y., Wang, Y., Zhang, K., Dong, W., Lv, B., & Bai, Y. (2015). Daily estimation of ground-level PM2.5 concentrations over Beijing using 3 km resolution MODIS AOD. *Environmental Science & Technology, 49*, 12280-12288

Yang, Q., Yuan, Q., Li, T., Shen, H., & Zhang, L. (2017). The Relationships between PM2.5 and Meteorological Factors in China: Seasonal and Regional Variations. *International Journal of Environmental Research and Public Health, 14*

You, W., Zang, Z., Pan, X., Zhang, L., & Chen, D. (2015). Estimating PM2.5 in Xi'an, China using aerosol optical depth: A comparison between the MODIS and MISR retrieval models. *Science of The Total Environment, 505*, 1156-116523